\documentclass[]{raa}
\usepackage{graphicx,times,subfig}
\usepackage{natbib}
\usepackage{amsmath}
\usepackage{cases}
\usepackage{amssymb}
\usepackage{longtable,textcomp}

\providecommand{\bm}[1]{\mbox{\boldmath$#1$\unboldmath}}

\begin{document}

   \title{Observational signature of a wind bubble environment for double neutron star mergers
}

   \volnopage{Vol.0 (201x) No.0, 000--000}      
   \setcounter{page}{1}           

   \author{Yong-Sen Li, Aming Chen,
   \and Yun-Wei Yu\mailto{}
%
      }

   \institute{Institute of Astrophysics, Central China Normal
University, Wuhan 430079, China \email{yuyw@mail.ccnu.edu.cn}\\
Key Laboratory of Quark and Lepton Physics (Central
China Normal University), Ministry of Education, Wuhan 430079,
China
          }

\date{Received~~2013 month day; accepted~~2013~~month day}

\abstract{During the in-spiral stage of a compact binary, a wind bubble could be blown into interstellar medium, if the electromagnetic radiation due to the binary orbital motion is strong enough. Therefore, short gamma-ray bursts (SGRBs) due to double neutron star mergers would in principle happen in a wind bubble environment, which can influence the propagation of the SGRB jet and the consequent afterglow emission. By calculating the dynamics and synchrotron radiation of the jet-driven external shock, we reveal that an abrupt jump could appear in the afterglow light curves of SGRBs and the observational time of the jump is dependent on the viewing angle. This light curve jump provides an observational signature to constrain the radius of the wind bubble and then the power of the binary electromagnetic radiation, by combining with gravitational wave detection.
}

   \authorrunning{Y. S. Li, A. Chen \& Y. W. Yu}            
   \titlerunning{Observational signature of a wind bubble environment for double neutron star mergers}  

   \maketitle

%
%
\section{Introduction}
Short-duration gamma-ray bursts (SGRBs) of $T_{90}\lesssim2$ s have been long hypothesized to originate from mergers of double neutron stars (NSs) or NS-black hole binaries (Paczy{\'n}ski 1986; Eichler 1989; Narayan et al. 1992). Since the first discovery of afterglow emission from SGRBs in 2005, this origin hypothesis has been increasingly supported by the large offsets of SGRBs
 from their host galaxies, by the non-detection of supernova associated with SGRBs, and by their event rates that can be connected with the cosmic star formation rates by power-law distributed time delays (Guetta \& Piran 2006; Nakar et al. 2006; Virgili et al. 2011;  Wanderman
\& Piran 2015).  On 17th August 2017, GRB 170817A was observed by the Fermi Gamma-Ray Telescope (Abbott et al. 2017a,b,c) from 1.7 s after the first detection of gravitational wave (GW) from a double NS merger by advanced LIGO. This first SGRB-GW association event eventually confirmed the long-hypothesized merger origin of SGRBs, although the very low luminosity of GRB 170817A (i.e., $L_{\rm iso}\sim10^{47}\rm erg~s^{-1}$) still makes it very different from typical SGRBs (generally $L_{\rm iso}\gtrsim10^{49}\rm erg~s^{-1}$).

The sites of compact binary mergers are usually far away from the center of their host galaxies. Therefore, the environment of a SGRB, where an external shock is driven by the SGRB jet, is widely considered to be low-density. As inferred from the fittings to the afterglow emission of SGRBs, the density range of their environmental medium is around $n\sim(10^{-3}-1)\rm cm^{-3}$ with a median value $\langle n\rangle<0.15\rm cm^{-3}$ (Berger 2014). In the afterglow fittings, an uniform interstellar medium (ISM) environment is usually assumed. However, this assumption is not always valid, in particular, if the pre-merger compact binaries can lose their orbital energy through intense electromagnetic radiation besides GW radiation (Medvedev \& Loeb 2013a). At the final stage of the in-spiral of compact binaries, in particular, double NS binaries, the electromagnetic radiation can in principle drive a relativistic binary wind to sweep up ISM by a shock wave. As a result, a nearly-isotropic wind bubble can be blown and expand continuously until a merger, which is bounded by a thin shell consisting of compressed ISM\footnote{This shocked ISM shell could be observed as a
faint radio source due to its synchrotron radiation (Medvedev \& Loeb 2013b).}. Therefore, compact binary mergers could happen at the center of a wind bubble. The resultant SGRB jets should first coast in a low-density wind nebula and then collide with a bubble shell, before the jets finally interact with the uniform ISM.

This paper is devoted to answer what observational signature can be caused by the interaction between a SGRB jet and a wind bubble, in particular, when the observation is off-axis and the jet has a complicated structure, just like for GRB 170817A. This work is somewhat similar to some previous studies for long GRBs, where a wind bubble is blown by the stellar wind from the progenitor Wolf-Rayet stars (Ramirez-Ruiz et al. 2001; Ramirez-Ruiz et al. 2005; Pe'er \& Wijers 2006; Kong et al. 2010). Our model is described in the next section. Results and discussion are given in Sections 3 and 4, respectively.

\section{The model}
\subsection{A binary wind bubble}
\begin{figure}
\centering
  \includegraphics[width=0.5\textwidth]{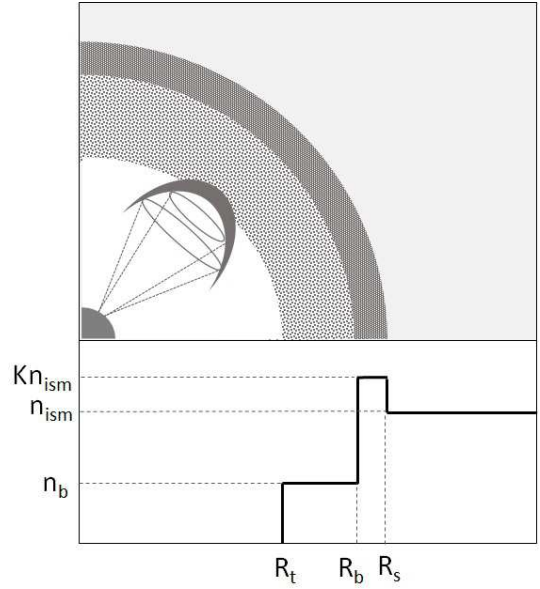}\\
  \caption{Illustration of a wind bubble environment for a double NS merger. The solid line represents the corresponding density profile of the bubble.}
  \label{fig: illustration}
\end{figure}

Besides GW radiation, in-spiraling NS binaries can also lose their orbital energy through electromagnetic radiation, due to the high orbital frequencies and the strong magnetic fields of the NSs. This energy release could be initially in the form of Poynting flux and gradually convert into an ultra-relativistic electron-positron wind, just like the formation and evolution of a pulsar wind (Medvedev \& Loeb 2013b). As a result, a wind bubble can be blown in the surrounding medium with a density profile of
\begin{equation}
n(r)=\left\{
\begin{array}{ll}
n_{\rm b}            &R_{\rm t}\leq r\leq R_{\rm b},\\
Kn_{\rm ism}  &R_{\rm b}<r<R_{\rm s}, \\
n_{\rm ism}    &R_{\rm s}\leq r,
\end{array}\right.
\end{equation}
where $n_{\rm b}$ and $n_{\rm ism}$ are the densities of the wind nebula and the un-shocked ISM, $K=(\hat{\gamma}+1)/(\hat{\gamma}-1)$ is given by the Rankine-Hugoniot jump condition with $\hat{\gamma}$ being the adiabatic index of the shocked material. The structure of a wind bubble is illustrated in Figure \ref{fig: illustration}. The characteristic radii $R_{\rm t}$, $R_{\rm b}$, and $R_{\rm s}$ are in principle functions of time, which are determined by the electromagnetic radiation process of the binary. However, a precise calculation of this electromagnetic radiation is unavailable, because of the unknown spins of the NSs and the unclear configuration of their common magnetosphere. In any case, for an amount of energy $E_{\rm s}$ that is primarily released during a period of $t_{\rm s}$, the outer radius of the shocked ISM shell can be estimated to
\begin{eqnarray}
R_{\rm{s}}&\sim&\left(\frac{3E_{\rm s}t_{\rm s}^2}{4\pi  n_{\rm ism}  m_{\rm p}}\right)^{1/5}=1.7\times10^{17}{\rm cm}\left(\frac{E_{\rm s,46}t_{\rm s,7}^2}{n_{\rm ism,-3}}\right)^{1/5}.
\end{eqnarray}
This is derived from the following equations: $M_{\rm sw}v_{\rm }^2\sim E_{\rm s}$, $R_{\rm s}\sim v_{\rm }t_{\rm s}$, and $M_{\rm sw}={(4\pi/3)}R_{\rm s}^3n_{\rm ism}m_{\rm p}$, where $M_{\rm sw}$ is the mass of the swept-up ISM in the shell and $v$ is the velocity of the external shock. Subsequently, the outer radius of the wind bubble $R_{\rm{b}}$ can be determined by $n_{\rm ism}R_{\rm s}^3=Kn_{\rm ism}(R_{\rm s}^3-R_{\rm b}^3)$ to
\begin{equation}
R_{\rm{b}}=\left({K-1\over K}\right)^{1/3}R_{\rm s}.
\end{equation}
Finally, by considering that the pulsar wind bubble nearly completely consists of electrons and positrons, its density can be given by
\begin{eqnarray}
n_{\rm{b}}&\sim&{3E_{\rm s}\over4\pi (R_{\rm b}^3-R_{\rm t}^3)\gamma_{\rm e^{\pm}}m_{\rm e}c^2}\nonumber\\&=&7.8\times10^{-6}[1-(R_{\rm t}/R_{\rm b})^{3}]^{-1}{\rm cm^{-3}}E_{\rm s,46}^{2/5}t_{\rm s,7}^{-6/5}n_{\rm ism,-3}^{3/5}\gamma_{\rm e^{\pm},5}^{-1},
\end{eqnarray}
where $\gamma_{\rm e^{\pm}}$ is the typical random Lorentz factor of $e^{\pm}$ in the shocked wind. By considering the mechanical balance between the two shocked regions, the radius of the termination shock of the wind can be determined by $R_{\rm t}=(E_{\rm s}/4\pi n_{\rm ism}v^2c t_{\rm s})^{1/2}$. In any case, the value of $R_{\rm t}$ would not substantively influence the dynamics of SGRB outflows. 



\subsection{A structured SGRB jet}
The relativistic outflows producing SGRBs are considered to be highly anisotropic and even collimated. The specific structure of these GRB jets is unclear and a simple ``top-hat" structure was usually adopted in literature (e.g., Lamb et al. 2005). However, very recently, observations to GRB 170817A showed that its afterglow light curves in all of the radio, optical, and X-ray bands share an identical behavior, i.e., continuously increasing from 2.3 days and reaching a peak at around 150 days (Margutti et al. 2017, 2018; Troja et al. 2017, 2018; D'Avanzo et al. 2018). This behavior cannot be explained by the ``top-hat" jet, but requires an angularly distributed jet that was observed off-axis (Lamb \& Kobayashi 2017; Mooley et al. 2018a,b; Granot et al. 2018; Lazzati et al. 2018). More specifically, the primary energy of the jet should be concentrated within a small cone of a half opening angle of $\theta_{\rm c}$. With an increasing angle relative to the jet axis, the energy density decreases gradually, accompanying by a decrease of Lorentz factor. Such a jet structure could be caused when the jet passes through and breaks out from a sub-relativistic intense outflow (i.e., the merger ejecta responsible for the kilonova emission). According to previous simulations of jet propagation and fittings to the afterglows of GRB 170817A, we adopt the following distributions for the kinetic energy and Lorentz factor of a SGRB jet (Dai \& Gou 2001; Zhang \& M{\'e}sz{\'a}ros 2002; Kumar \& Granot 2003):
\begin{equation}
{\varepsilon(\theta)\equiv \frac{dE_{\rm jet}}{d\Omega}=}\left\{
\begin{array}{ll}
\varepsilon_{\rm c}  ,   &\theta\leq \theta_{\rm c}, \\
\varepsilon_{\rm c}\left(\frac{\theta}{\theta_{ \rm c}}\right)^{-k} ,  &\theta_{\rm c}<\theta<\theta_{\rm m},
\end{array}\right.\label{eq: Edis}
\end{equation}
and
\begin{equation}
{\Gamma_{\rm 0}(\theta)=}\left\{
\begin{array}{ll}
\eta   ,  &\theta\leq \theta_{\rm c},\\
\eta\left(\frac{\theta}{\theta_{\rm c}}\right)^{-k}+1 ,  &\theta_{\rm c}<\theta<\theta_{\rm m},
\end{array}\right.\label{eq: Lordis}
\end{equation}
where $\theta_{\rm m}$ is defined as the maximum angle of the jet. The index $k$ is a constant that can be deduced from the luminosity distribution of local event rate density (Pescalli et al. 2015; Xiao et al. 2017). The interaction of such a structured jet with a pre-merger wind bubble is on the focus of this paper.

\subsection{Dynamics of jet external shock}
\begin{figure}
\centering
  \includegraphics[width=0.65\textwidth]{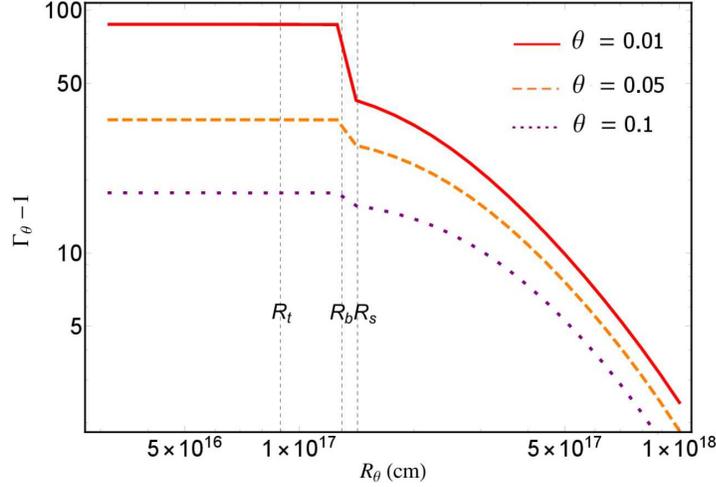}\\
  \caption{Lorentz factors as functions of $r_{\theta}$ for different jet rings. The model parameters are taken as $\theta_{\rm c}=0.02$, $\theta_{\rm m}=0.1$, $k=1$, $\varepsilon_{\rm c}=10^{50.5}~\rm erg ~sr^{-1}$, $\eta=88$, $n_{\rm ism}=0.01$, $E_{\rm s}=4.1\times10^{46}$ erg, $R_{\rm s}\simeq1.42\times10^{17}$ cm,  and $R_{\rm b}\simeq1.29\times10^{17}$ cm. }
  \label{fig: dyn}
\end{figure}

In order to calculate the dynamical evolution of a structured jet, we separate the jet into a series of differential rings. The energy per solid angle and the Lorentz factor of these rings are given by according to Eqs. (\ref{eq: Edis}) and (\ref{eq: Lordis}). For simplicity, the dynamical evolution of the rings is considered to be independent with each other by ignoring their possible lateral expansion/motion. Then, the following equation can be used (Huang et al. 2000):
\begin{equation}
\frac{d\Gamma_{\theta}}{d\Sigma_{\rm sw,\theta}}=-\frac{\Gamma_{\theta}^{2}-1}{\Sigma_{\rm ej,\theta}+2\Gamma_{\theta} \Sigma_{\rm sw,\theta}},
\end{equation}
where $\Sigma_{\rm ej,\theta}=dM_{\rm ej}/d\Omega=\varepsilon_{\theta}/\Gamma_{\theta,0}$ is the jet mass per solid angle at angle $\theta$ and the corresponding swept-up ISM mass is determined by
\begin{equation}
\frac{d\Sigma_{\rm sw,\theta}}{dr_{\theta}}=n m_{\rm p}r_{\theta}^{2},
\end{equation}
where $r_{\theta}$ is the radius of the external shock front driven by the propagation of the SGRB jet.
Numerical results of the dynamical calculations are presented in Figure \ref{fig: dyn} for different differential rings of the jet. The sharp decay of the Lorentz factors at $r_{\theta}=R_{\rm b}$ is due to the collision of the jet rings with the compressed bubble shell.

\begin{figure}
\centering
  \includegraphics[width=0.4\textwidth]{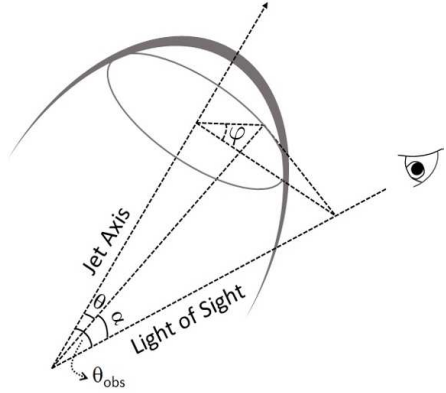}\\
  \caption{The relation among the angles $\theta$, $\varphi$, $\theta_{\rm obs}$, and $\alpha$ for a differential swept-up ISM element.}
  \label{fig: angles}
\end{figure}
\subsection{Shock synchrotron emission}
For an differential element of a mass $\Sigma_{\rm sw}(r,\theta,\varphi)d\varphi d\theta$, the synchrotron luminosity contributed by electrons in this mass can be calculated by an analytical method delivered by Sari \& Piran (1998) as
\begin{equation}
\ell'_{\nu'}(r,\theta,\varphi)d\varphi d\theta={\Sigma_{\rm sw}(r,\theta,\varphi)d\varphi d\theta\over m_{\rm p}}{m_ec^2\sigma_{\rm T}B'(r,\theta,\varphi)\over 3e}S'(\nu'),
\end{equation}
where the superscript prime indicates the quantities are measured in the comoving frame of shocked region, $B'(r,\theta,\varphi)$ is the magnetic field strength, and $S'(\nu')$ represents the dimensionless synchrotron spectrum. This spectrum can be approximately expressed by a multi-broken power law as
\begin{equation}
S'(\nu')=\left\{
\begin{array}{ll}
(\nu^{'}/\nu^{'}_{\rm l})^{1/3},  &\nu^{'}\leq\nu^{'}_{l} ,   \\
(\nu^{'} / \nu^{'}_{\rm l})^{-(q-1)/2}    ,   &\nu^{'}_{\rm l}<\nu^{'}<\nu^{'}_{\rm h} ,     \\
(\nu^{'}_{\rm h}/\nu^{'}_{\rm l})^{-(q-1)/2}~(\nu^{'}/\nu^{'}_{\rm h})^{-p/2}  , &\nu^{'}_{\rm h}\leq\nu^{'} ,
\end{array}\right.\label{eq: Snu}
\end{equation}
where the broken frequencies $\nu'_{l}(r,\theta,\varphi)$ and $\nu'_{h}(r,\theta,\varphi)$ are determined by the energy distribution of electrons, $p$ is the power-law index of shock-accelerated electrons, and $q=2$ or $q=p$ for the rapid and slow cooling cases, respectively. Then, the observed flux at an observational time $t$ can be obtained by integrating over the whole solid angle of the jet as
\begin{equation}
F_{\rm \nu}(t)={1\over 4\pi D^{2}_{\rm L}}\int_{0}^{\theta_{\rm m}}\int_{0}^{2\pi} \frac{\ell^{'}_{\nu^{'}}(r,\theta,\varphi)}{\Gamma_{\theta}^3(1-\beta_{\theta}\cos\alpha)^3}d\varphi d\theta
\end{equation}
where $D_{\rm L}$ is the luminosity distance of the SGRB, $\theta_{\rm obs}$ is the viewing angle with respect to the jet axis, and the angle $\alpha$ which is defined between the emitting differential element and the light of sight can be determined by
\begin{eqnarray}
\cos\alpha  =
\frac{\cos\theta_{\rm obs}}{2\cos\theta}\left(1+\frac{\cos^{2}\theta}{\cos^{2}\theta_{\rm obs}}-\sin^{2}\theta-\cos^{2}\theta\tan^{2}\theta_{\rm obs}+2\cos\theta\sin\theta\tan\theta_{\rm obs}\cos\varphi\right).
\end{eqnarray}
The relation among the angles $\theta$, $\varphi$, $\theta_{\rm obs}$, and $\alpha$ is showed in Figure \ref{fig: angles}. Finally, the radius $r$ of emitting material can be connected with the observational time by $r=ct/(1-\beta_{\theta}\cos\alpha)$, where $\beta_{\theta}=(1-\Gamma_{\theta}^{-2})^{1/2}$.

\begin{figure}
\centering
  \includegraphics[width=0.6\textwidth]{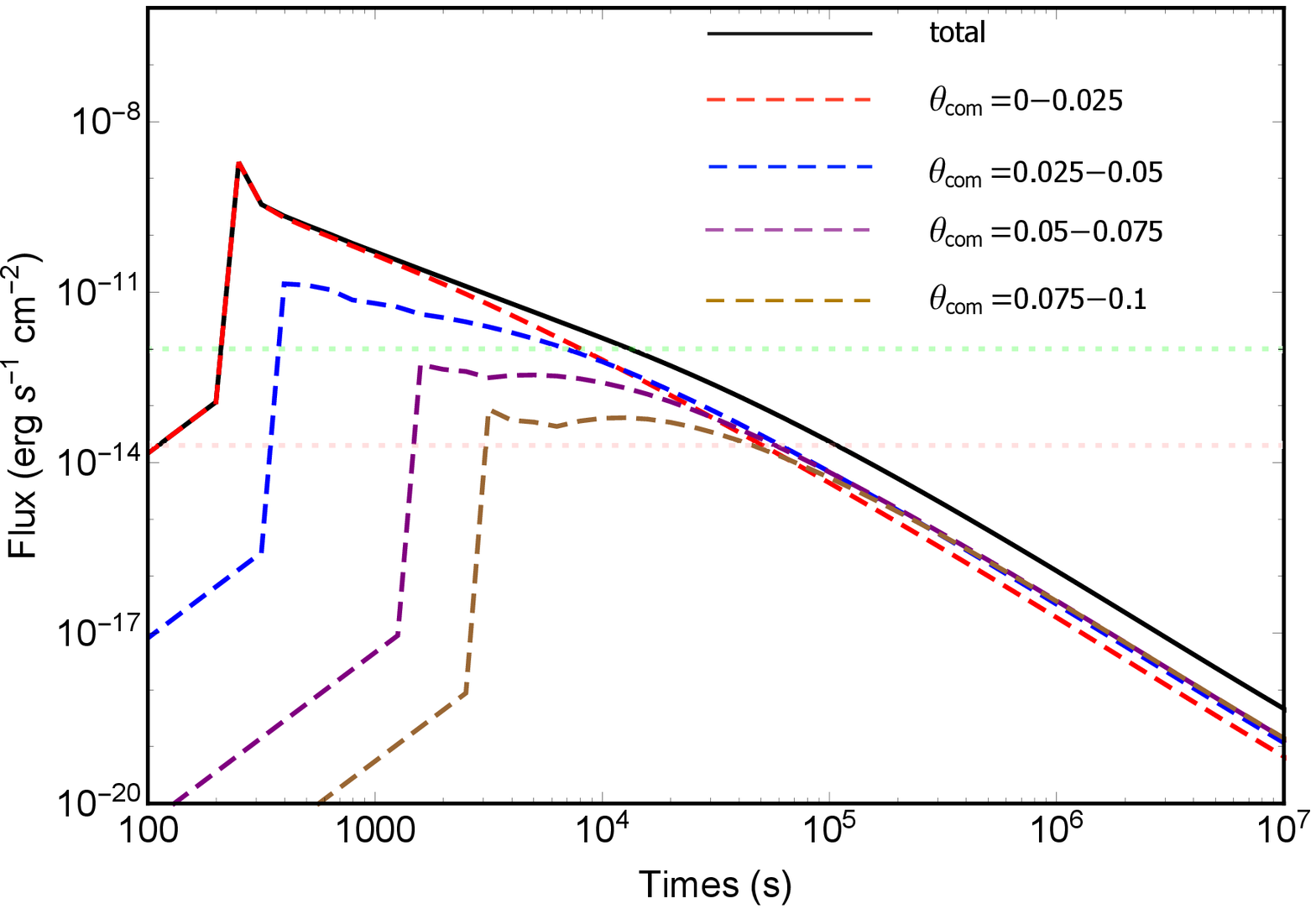}\\
  \caption{An X-ray (1 keV) afterglow light curve (solid line) arising from a SGRB jet interacting with a binary wind bubble. The dashed lines present the contributions of the material within the $\theta$ ranges as labeled. The dotted lines present the sensitivities of Swift XRT and Einstein-Probe (EP) in $10^{4}$ seconds.}
  \label{fig: lihgtcurve1}
\centering
  \includegraphics[width=0.6\textwidth]{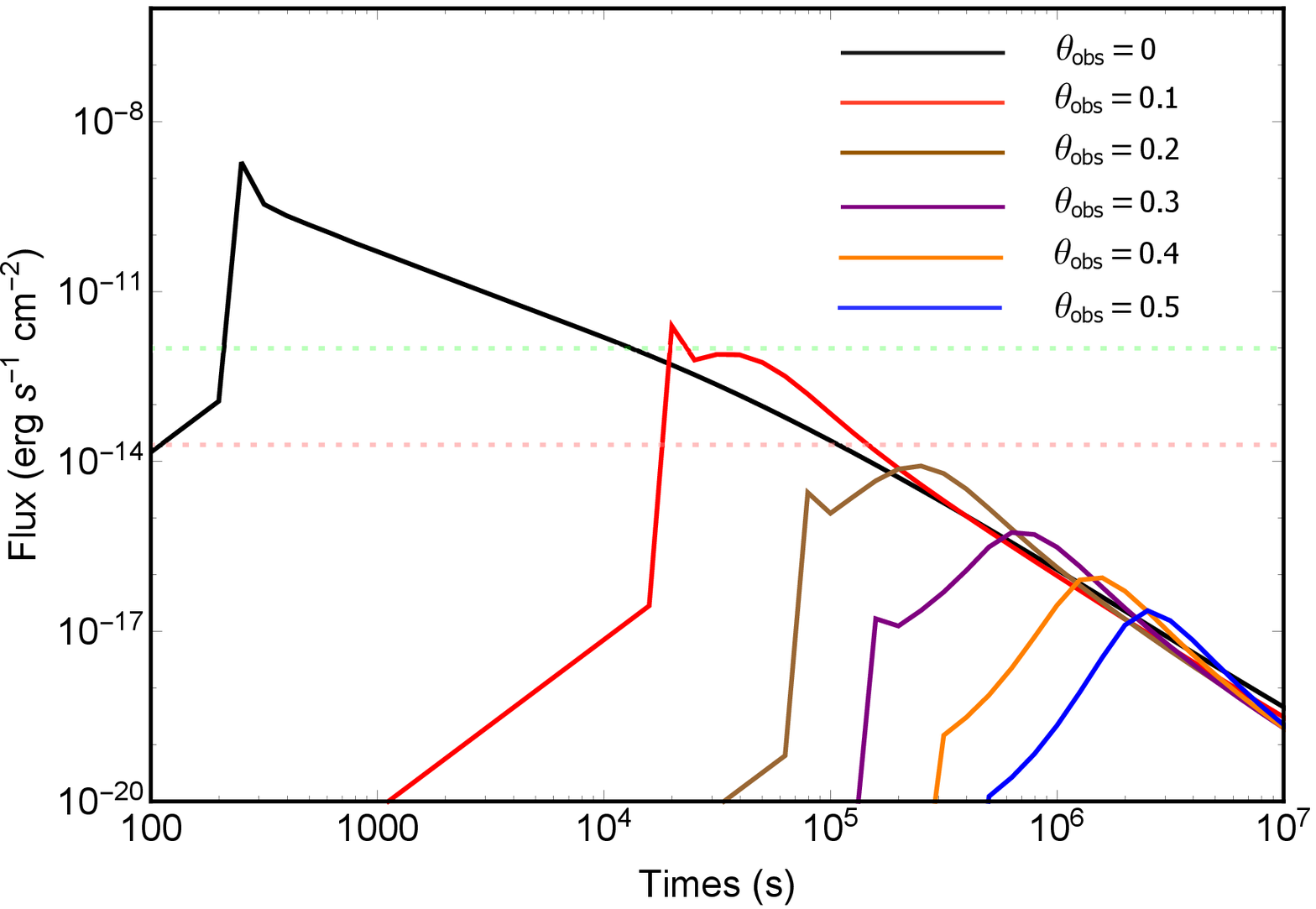}\\
  \caption{X-ray (1 keV) afterglow light curves for different viewing angles as labeled.}
  \label{fig: lihgtcurve2}
  \centering
  \includegraphics[width=0.6\textwidth]{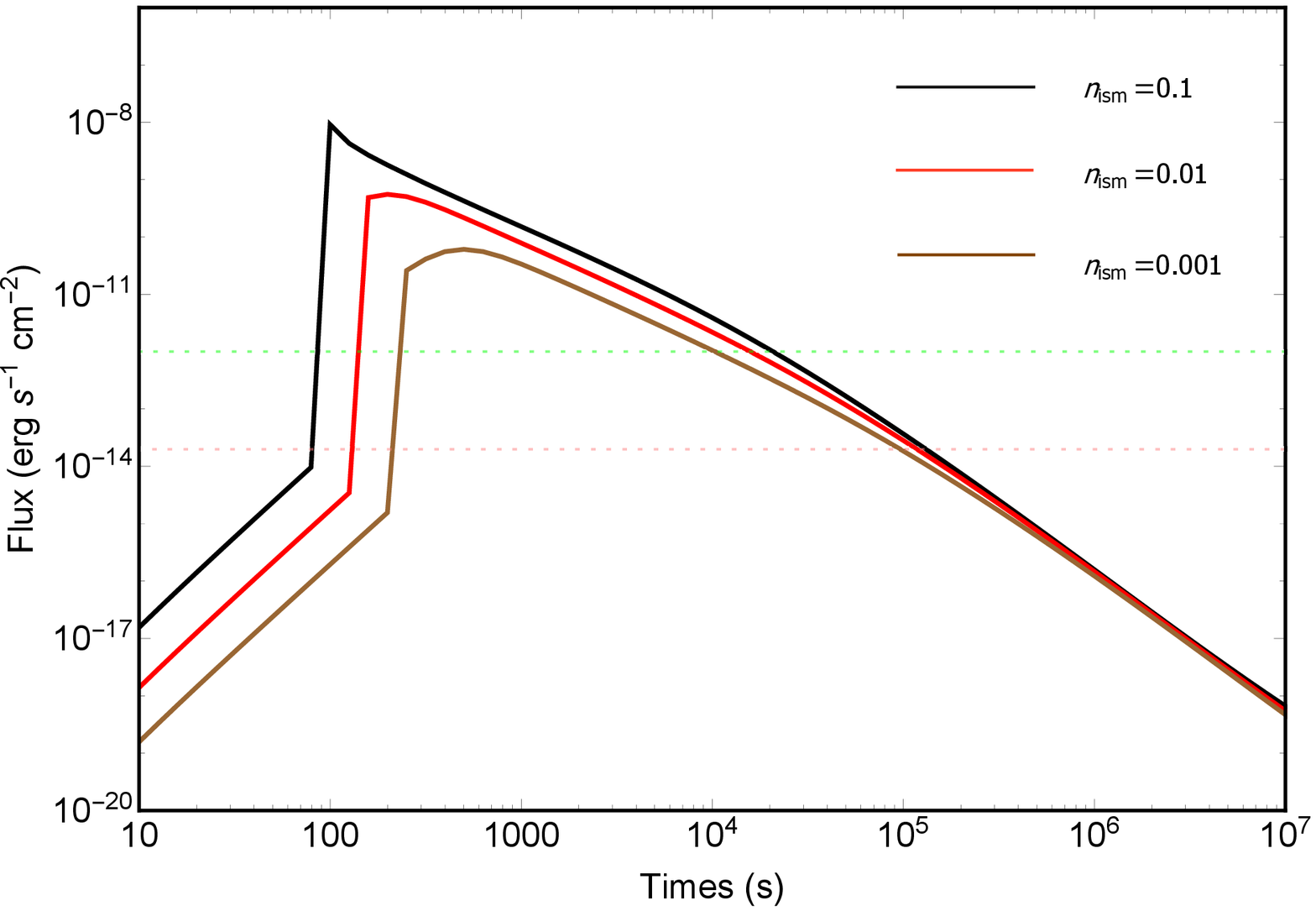}\\
  \caption{X-ray (1 keV) afterglow light curves for different environment medium densities.}
  \label{fig: lihgtcurve3}
\end{figure}

\section{Results}
In Figure \ref{fig: lihgtcurve1}, we present an example afterglow light curve observed on-axis, where an X-ray frequency is taken. The contributions from different differential rings are represented by the dashed lines. As the early afterglow emission is dominated by the contribution from the core of the jet, the emission from large angles would become more and more important at late times. In comparison with previous afterglow calculations, an abrupt jump appears in our light curve at a time $t_{\rm jp}$, which arises from the abrupt jump of the environmental density at $R_{\rm b}$. For an on-axis observation, we have $t_{\rm jp}\approx R_{\rm b}/(2\eta^2c)$. Therefore, this light curve jump can be regarded as an observational signature of the wind bubble environment. Furthermore, since the velocity of the jet material decreases with increasing $\theta$, different jet rings collide with the bubble shell at different time. Therefore, for off-axis observations, the light curve jump time would be delayed. Specifically, we have  $t_{\rm jp}\approx R_{\rm b}/[c(1-\beta_{\theta_{\rm obs}})]$. Such off-axis observation and environment density effects are presented in Figure \ref{fig: lihgtcurve2} and Figure \ref{fig: lihgtcurve3}.


\section{Discussion}
As the confirmation of the origin of SGRBs from double NS mergers by the GW 170817-GRB 170817A association, it has become a frontier topic to answer how the mergers happen and how GRB outflow forms and evolves. In particular, the electromagnetic radiation by in-spiraling NS binaries is still completely unknown, which however can influence the orbital decay of the binaries and modify the environments where SGRBs occur. By supposing a wind bubble driven by the binary electromagnetic radiation, we calculate the dynamics and synchrotron radiation of an external shock arising from the interaction between a structured SGRB jet and the bubble. As a result, it is revealed that an abrupt jump could appear in the afterglow light curves of SGRBs and the observational time of the jump is dependent on the viewing angle. Therefore, the discovery of this light curve jump can be used to constrain the radius of the wind bubble and then the power of the binary electromagnetic radiation, after the viewing angle has been fixed by the peak time of the afterglow and by the GW detection.

\begin{acknowledgements}
This work
is supported by the National Natural Science Foundation of China
(Grant No. 11822302 and 11833003) and the
self-determined research funds of CCNU from the colleges'
basic research and operation of MOE of China.
\end{acknowledgements}

\label{lastpage}

\end{document}